# Measurements on a little known sound source - the Vortex Whistle

Ulf Kristiansen[1], Muriel Amielh[2],

[1]Acoustics Research Centre, Norwegian University of Science and Technology, Trondheim, Norway
[2]Aix Marseille Université, CNRS, Centrale Marseille, IRPHE UMR 7342, Marseille, France
Contact email: ulf.kristiansen@ntnu.no

## Abstract

Acoustic measurements on a vortex whistle corroborates earlier findings that the frequency increases linearly with the velocity of the air flowing into the whistle. Measurements in a reverberant chamber shows that the acoustic power generated by the whistle increases close to the sixth power of the frequency. PIV and hot-wire measurements give quantitative information on the flow. It is shown that the flow exits the pipe as a swirling annulus with a rotation corresponding to the frequency of the sound, and that the center of the swirl follows a nearly circular path in the exit plane. The PIV measurements also indicate a toroidal vortex with inflow along the center line close to the exit of the whistle.

## 1  Introduction

The main characteristic of a vortex whistle is that it produces a pure tone sound proportional to the amount of fluid flowing through it. The normal design is such that an airflow enters a circular volume tangentially, spirals up, and exits with an increased rotation rate through a cylinder of narrower bore, see Figure 1. The whistle was first described in the literature by B. Vonnegut in 1954 [1]. He found that the whistling frequency is linearly proportional to the flow rate through the system and that it works equally well in water as in air. Visualizing the phenomenon using colored dyes in water, Watanabe and Sato [2] showed that during whistling, part of the fluid in the exit cylinder is stationary and likens this to the neck mass in an Helmholtz resonator. It was also observed that the fluid flows out of the exit opening not as an axisymmetric jet, but rather as a spiraling movement along the exit cylinder's wall. Theoretical perturbation analyses of two dimensional and rotating flow equations by respectively Michelson [3] and R. Chanaud [4], suggested that the whistling frequency equals the flow rotation frequency, or multiples of this. The sound frequency being given by the rotation frequency was later verified experimentally by Sato and Watanabe [5]. Further water observations on the exiting flow, i.e. that also the axis of the swirling outflow precess and that a backflow in the form of a toroidal vortex exists in the center was made by R. Chanaud [6]. The same author measured the sound's directivity to be similar to that of a dipole with its axis in the plane of the tube exit. A rotating dipole was therefore assumed to be the principal sound source [4].

The vortex whistle has found a practical application in flow metering devices. Details of such are found in the works by Sato and Watanabe [5] and Yan-bin Di et al. [7]. Other applications could be as underwater sound sources and possible musical instruments.

The purpose of the research presented in this paper has been to investigate the vortex whistle in more detail, by acoustic and fluid dynamic experiments. In section 2 we present acoustic measurements on the whistling frequency vs flow rate and verify the earlier findings that linear relationships exist. We also present results from an experiment where the acoustic power has been measured at different running conditions. The acoustic power was found to increase with the $6^{th}$ power of the frequency. Section 3 presents flow velocity measurements. These were done using a x-type hot-wire for temporal / spectral analysis and Particle Image Velocimetry (PIV) measurements for spatial analysis of the flow field. PIV measurements in the exit plane allowed us to obtain details of the mean rotation and instantaneous flow fields. The precession movement was made clear, as well as the temporal location of the center of the instantaneous vortex and high velocity parts of the flow. Section 4 gives a summary of the investigation. The flow behavior has earlier mostly been discussed quantitatively by observations in water using colored dyes. Our present technique allows quantitative flow measurements with the whistle operating in air.

## 1.1 Dimensions of the whistle

The whistle was made in glass by a glass-blower, see figure 1 The inlet and outlet pipes are connected to a cylindrical volume of $241\,\text{m}^3$ ($54\,\text{mm}$ diameter).The inner pipe diameters are $14\,\text{mm}$ and $19\,\text{mm}$ for the inlet and outlet pipes respectively. The length of the outlet pipe is $65\,\text{mm}$. The center of the inlet tube is displaced $29\,\text{mm}$ from the bottom. Tests by Chanaud [4] showed that such a displacement had no influence on the emitted sound. The glass thickness in the outlet tube is $2\,\text{mm}$. A small entry at the bottom (inside diameter $2.7\,\text{mm}$) serves two purposes. It allows measurements of the static pressure at the bottom of the flask, and for the PIV measurements to be discussed in section 3, it allows the introduction of small oil droplets. During most acoustics measurements this opening was closed by a piece of modeling clay. The inlet pipe was connected to a compressed air system having a reduction valve during the experiments.

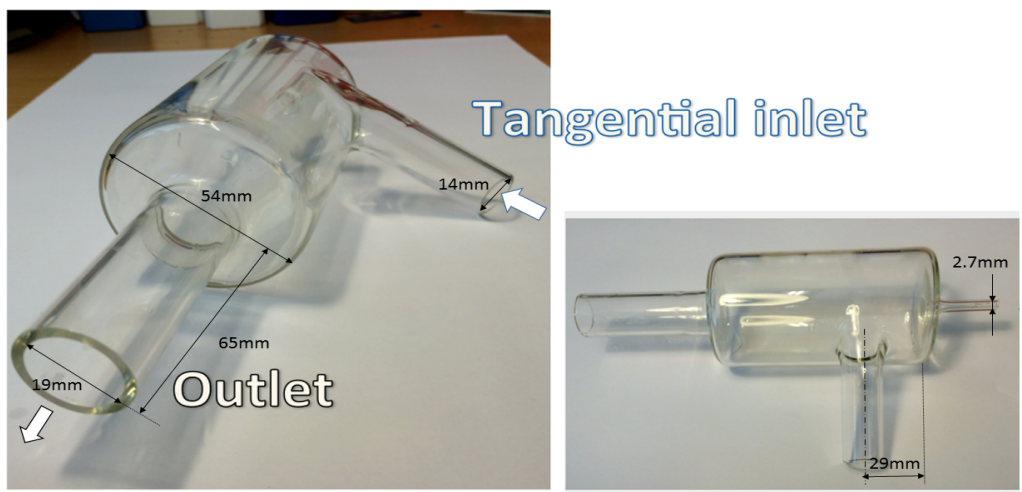

Figure 1: Pictures of the flask

## 2 Acoustic measurements

### 2.1 Frequency *vs.* velocity measured in inlet pipe

To verify that our whistle behaved like the ones investigated by previous investigators, the principal frequency was measured as function of the mean inflow velocity. The whistle was fed by a compressed air system of 8 bars. The amount of air was controlled by means of a reduction valve. The whistling frequency was measured using a narrow band (frequency resolution set at 1Hz) analyzer (faberacoustical Signalscope 6.1.1 implemented on an iPhone 5S which was equiped with a EDUTIGE EIM-003 microphone). The sampling rate was set at $48\,\text{kHz}$. The air velocity was measured by a $3\,\text{mm}$ outer diameter Pitot tube measuring at the center line of a $18\,\text{mm}$ inner diameter elastic hose connecting the air supply to the flask. The pressure tappings on Pitot tube were connected to a ALNOR AXD540 pressure meter. Figure 2 shows a typical spectrum.

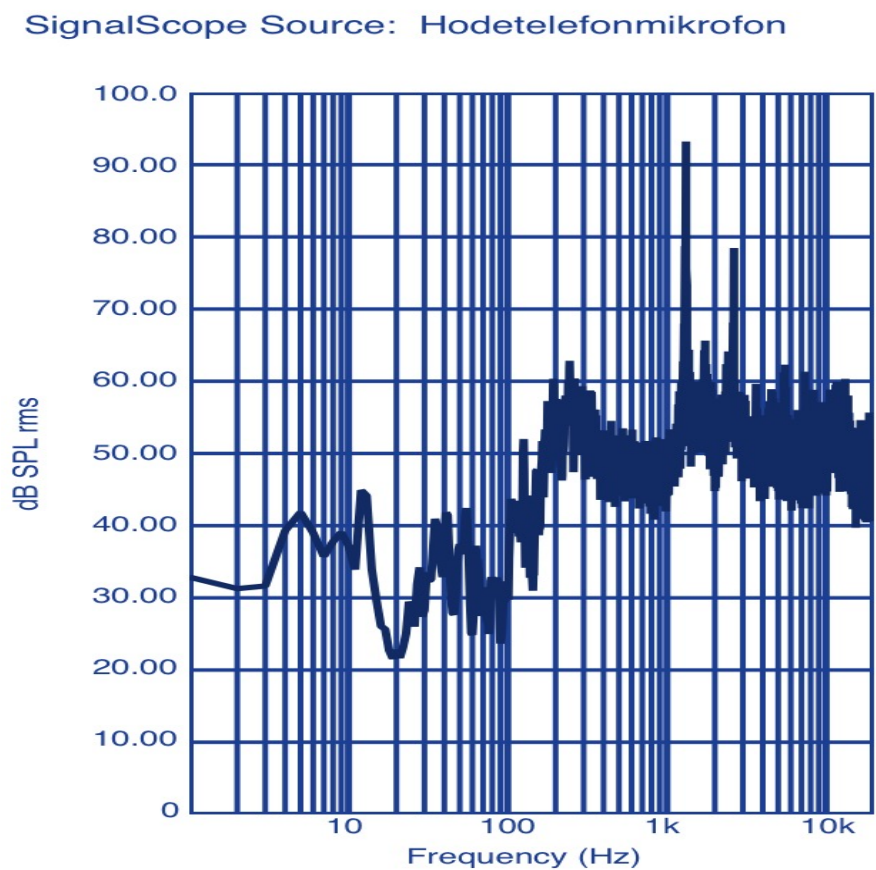

Figure 2: Example of measured frequency spectrum.

Figure 3 shows the frequency as a function of the area adjusted center velocity for the $14\,\text{mm}$ inlet pipe .

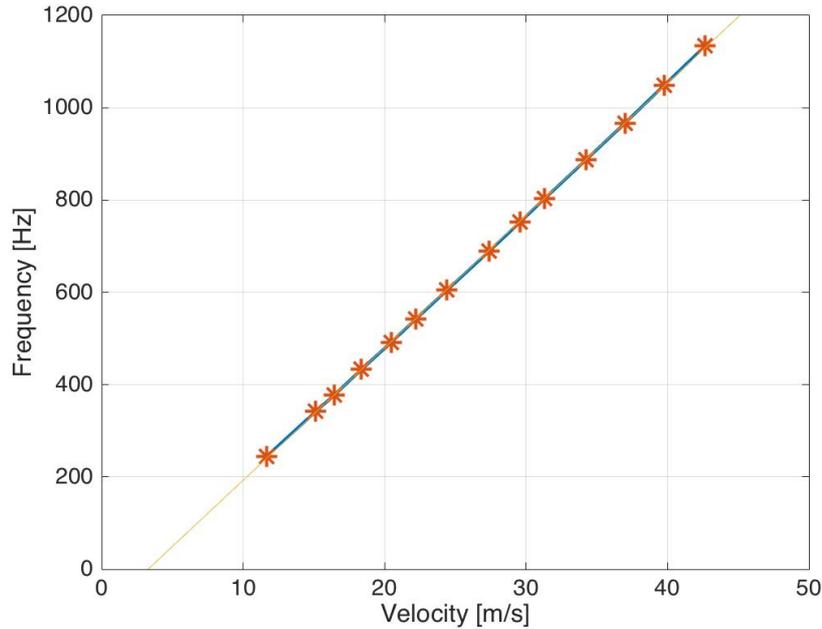

Figure 3: Frequency vs. center line velocity in inlet duct

The experimental regression line shows

$$f = 28\,U_{inlet} - 110 \tag{1}$$

for the frequency vs. center line inlet velocity. At low velocities *i.e.* $U_{inlet} < 12\,\mathrm{m\,s^{-1}}$ it was not possible to detect a distinct peak. It is also seen that the regression line does not go through the origin, but that a certain velocity, $U_0$ is needed before a given velocity increment will give a constant frequency shift. By opening the small tube at the bottom of the flask to the surroundings it was observed that the frequency would decrease by close to 8%. At the same time it was observed that air was sucked into the flask through this opening.

## 2.2 Sound power measurement

The whistle was positioned in a reverberant room having an outlet for pressurized air. The sound pressure was measured by moving a handheld Norsonic 140 sound level meter around in the room (random walk) using an integration time of 59 seconds and 1/3 octave filtering. The whistling frequencies, which appear as pure tones, were adjusted (by controlling the air flow) to be close to the center frequencies of the 1/3 octave bands. The levels were calibrated by first measuring the output of a reference sound source (Bruel & Kjaer Type 4204) using the same measuring procedure.

An interest in the present experiment would be to find the generated acoustic power as a function inlet velocity. The velocity exponent is indicative of the type of source we are dealing with. For the present reverberant chamber set-up it was however not practical to measure the inlet velocity directly. We could however measure the dominant frequency, and as we had found this to be linearly proportional to the velocity we sought a relationship

$$P = a f^x. \tag{2}$$

Expressing this by power level, referenced to $P_0 = 10^{-12}$ Watt, we get

$$L_W = 10lg(\frac{P}{P_0}) = 10lg(\frac{a}{P_0}) + x \cdot 10lg(f). \tag{3}$$

The result is plotted in figure 4 below. The linear regression line shows the exponent $x$ to be very close to 6. This, and the fact that the frequency varies linearly with velocity in the active region strongly suggests the source to be of the dipole type, [8]. Using the regression line

$$L_W = 6 \cdot 10lg(f) - 95, \tag{4}$$

the sound power emitted from the whistle can be written

$$P \approx 3.2 \cdot 10^{-22} f^6 \quad \text{Watt}. \tag{5}$$

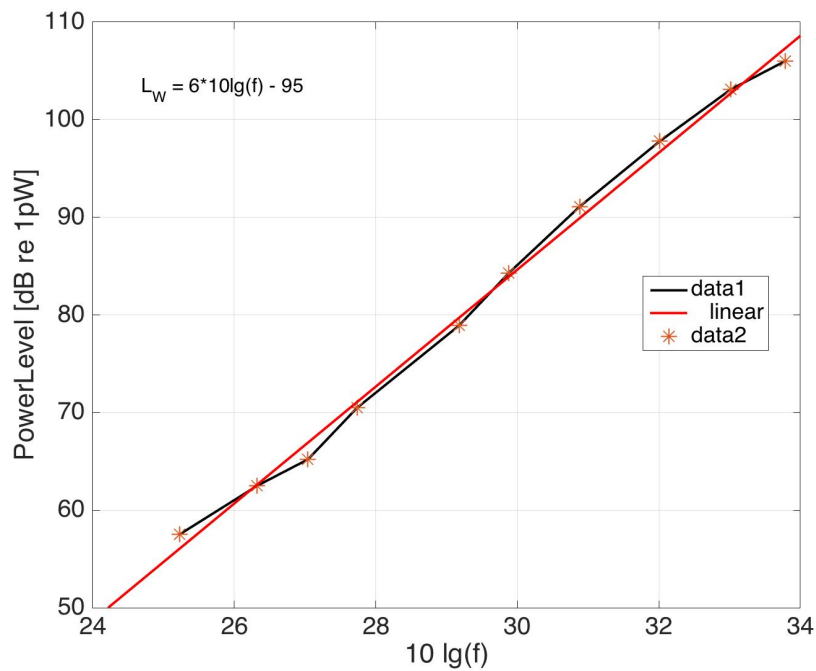

Figure 4: Sound power level vs. $10\lg(f)$

Comparing the linear approximation with the measured values in figure 4, we find that the norm of the residuals for the 10 measuring points is 3.27dB.

## 3 Flow measurements

Hot-wire and Particle Image Velocimetry (PIV) measurements were performed in the three planes shown in figure 5. Hot-wire measurements were limited to the $x = 0$ plane. The interest of these measurements by hot-wire is the high frequency response of the sensor allowing wave forms to be obtained and spectral analysis to be performed on the velocity signals. These temporal and spectral performances obtained by the hot-wire technique are completed by the spatial investigations provided by PIV measurements. The positioning of the planes allowed measurements of the axial (U), radial (V), and azimuthal (W) velocity components indicated in figure 6. The measurements were limited to relatively low (15 - 20 m/s) inlet velocities.

Proceedings of the 39th Scandinavian Symposium on Physical Acoustics, Geilo, Norway, Jan. 31 – Feb. 3, 2016

## 3.1 Set-up for PIV measurements

Details of the experimental set up are listed in tables 1 and 2.

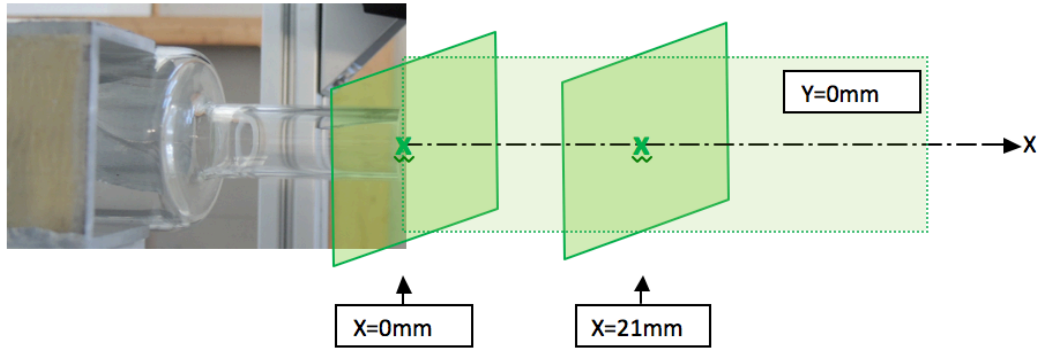

Figure 5: Sketch showing the PIV measuring planes

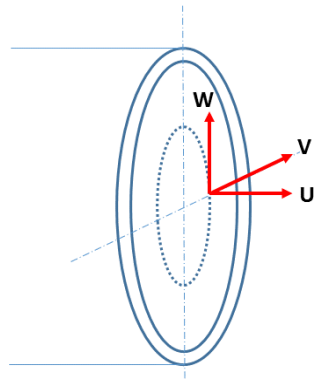

Figure 6: Sketch defining the axial (U), radial (V), and azimuthal (W) velocity components at a point in the exit (x=$0\,\text{mm}$) plane.

| PIV details | $\Delta t_{piv}$ | Scale | Field of view dimension | PIV fields number | Measured components |
|---|---|---|---|---|---|
| $X = 0\,\text{mm}$ | 25 $\mu$s | 24.2/922 mm/px | 26.5 x 26.5 $\text{mm}^2$ | 499 | V and W |
| $X = 21\,\text{mm}$ | 25 $\mu$s | 17/678 mm/px | 25.3 x 25.3 $\text{mm}^2$ | 456 | V and W |
| $Y = 0\,\text{mm}$ | 45 $\mu$s | 24.2/286 mm/px | 85.3 x 85.3 $\text{mm}^2$ | 499 | U and V |

Table 1: Experimental conditions

| | |
|---|---|
| Camera: | KODAK ES 1.0 1008x1018px |
| Pulsed laser Nd: | Yag, $532\,\mathrm{nm}$, $200\,\mathrm{mJ}$, $10\,\mathrm{Hz}$, Spectra-Physics |
| Image acquisition: | Video Savant software |
| Triggering by: | Stanford synchroniser |
| Image postprocessing: | DynamicStudio (Dantec Dynamics) |
| Seeding: | aerosol of micronic olive oil droplets |
| PIV fields postprocessing: | Matlab |

Table 2: List of equipments used during flow measurements

### 3.2 Set-up for hot-wire measurements

Hot-wire probe used for U and W velocity components measurements is constituted of an X-array of two hot- wires of $5\,\mu$m diameter and $1.25\,\mathrm{mm}$ length, separated by $1\,\mathrm{mm}$ (55P61 probe, DANTEC Dynamics). The sensor plane is parallel to probe axis. The probe is calibrated in velocity (range $0.2$ - $60\,\mathrm{m\,s^{-1}}$) and angle (range $-30\,°$ - $+30\,°$). The X-hot-wire is controlled by constant temperature anemometer Streamline unit (DANTEC Dynamics) and data acquisitions are driven by the associated software Streamware. For each position of the velocity profile, 262 144 ($2^{18}$) samples are acquired at a $50\,\mathrm{kHz}$ sampling frequency. The functioning principle of hot-wire does not permit to distinguish $U > 0$ from $U < 0$ so that the hot-wire measurement in the central part of the $X = 0\,\mathrm{mm}$ section is questionable because the flow is mainly reverse in this region. X-hot-wire measurements are only punctual, that is why the probe is displaced by manual 3D-micromanipulators in order to explore a radial velocity profile at the bottle exit (section $X = 0\,\mathrm{mm}$).

### 3.3 Results: mean flow measurements

From the PIV measurements, the mean velocity magnitude and corresponding vectors are calculated and shown as $(V^2 + W^2)^{1/2}$ in the $X = 0$ plane, figure 7, and $(V^2 + U^2)^{1/2}$ in the $Y = 0$ plane in figure 8.

In the $X = 0$ plane, the counterclockwise rotation is clearly seen in the figure 7. For a whistling frequency of $397\,\mathrm{Hz}$, the maximum mean velocity in the azimuthal direction was found to be $W = 11\,\mathrm{m\,s^{-1}}$ at $76\%$ of the pipe radius. Figure 9 shows the mean velocity components $V$ and $W$ extracted along the diameters $Y_{piv} = 0$ and $X_{piv} = 0$ in figure 7. A x-type hot wire was also positioned to measure the $U$ and $W$ velocity components, and a measurement of the $W$ component is included in the figure. The set up depends on a positive value of the outflow ($U$) component, and the discrepancy between the hot-wire and PIV results towards the middle of the pipe is probably due to the fact that the flow in this region is in fact inwards, see figure 8.

The mean velocity magnitude and vectors in the $Y = 0$ plane, figure 8, again shows the flow leaving the pipe as a rotating annulus. A slight asymmetry in the up and down directions was ascribed non ideal laboratory conditions. A striking feature of the flow is the core of reverse flow seen in the middle, suggesting a toroidal type vortex. Figure 10 shows the reversed flow to be present up to around $30\,\mathrm{cm}$ or around 15 outlet diameters. The non zero mean axial $V$ component is again accredited the unevenness of the flow.

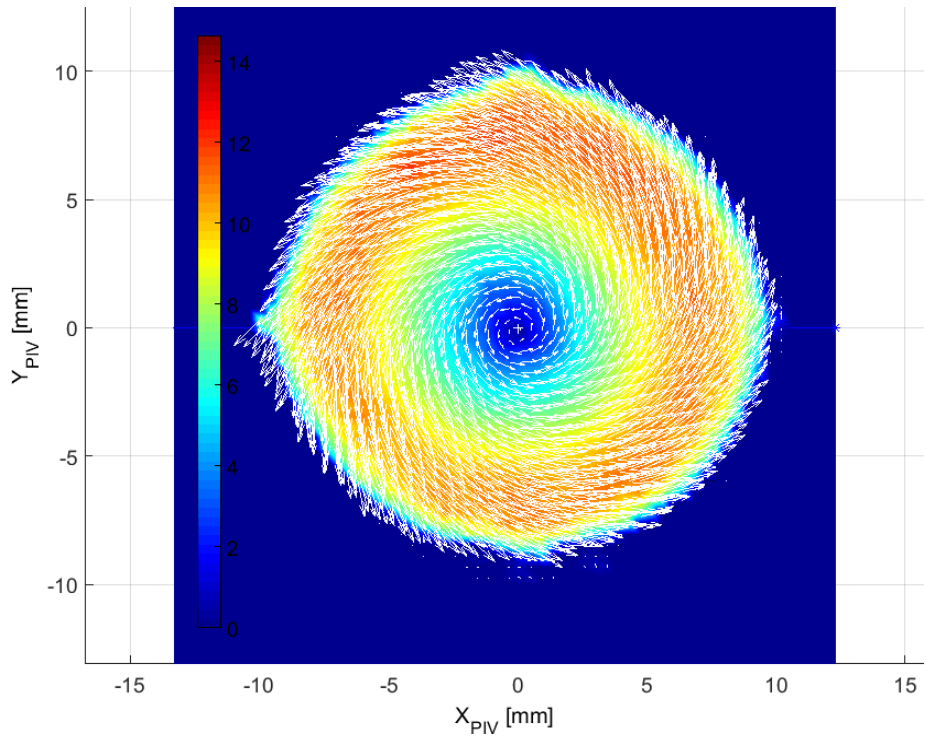

Figure 7: Mean velocity magnitude, $(V^2 + W^2)^{1/2}$, and vectors at $X = 0$.

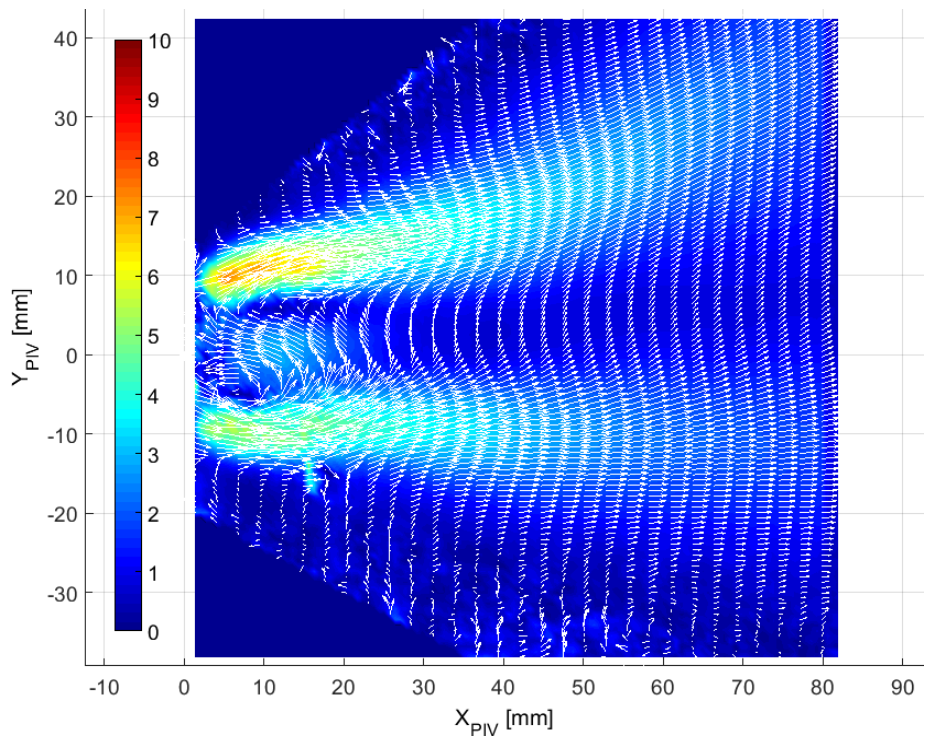

Figure 8: Mean velocity magnitude, $(U^2 + V^2)^{1/2}$ in the $Y = 0$ plane (from PIV).

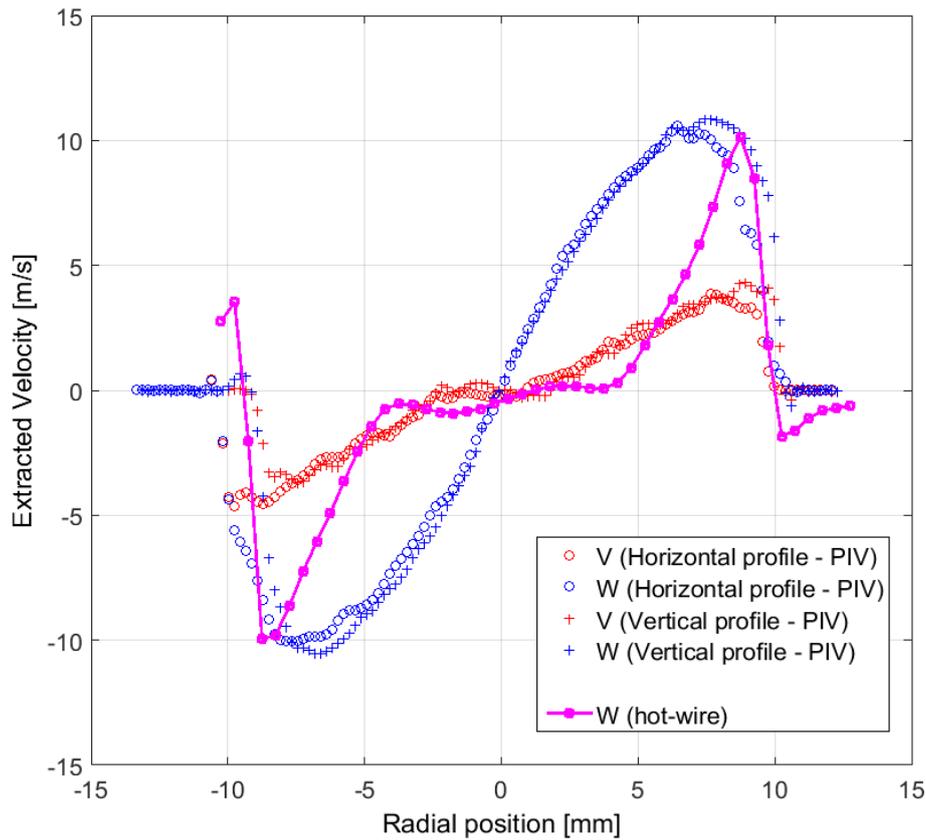

Figure 9: Radial profiles of mean velocity components: $V$ and $W$. PIV data extracted from the lines $Y_{piv} = 0$ in figure 8.

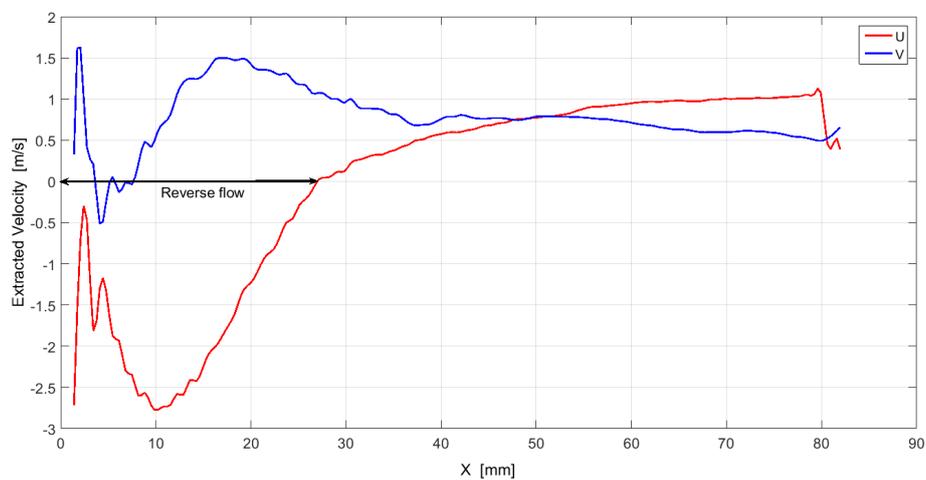

Figure 10: Axial profiles of $U$ and $V$ components. PIV data extracted from the lines $Y_{piv} = 0$ and $X_{piv} = 0$ in figure 7.

### 3.4  Results: instantaneous velocity values

Studying the PIV "snapshots" it becomes clear that the exiting flow is not as symmetric as the mean velocity maps suggest. The center of the instantaneous vortex is seen to follow a nearly circular path to give the outflow a precessional movement. An assembly of vortex central points is shown in figure 11. The indicated circle is slightly displaced from the geometric center of the outflow pipe. Table 3 shows four PIV "snapshots" taken at different times during a cycle. It is seen that the high velocities are roughly located in the region in between the center of the vortex and the outlet pipe wall.

The hot-wire probe was positioned such that the $U$ and $W$ components could be separated. Displaying the time signals recorded by the hot-wire, figure 12, shows more clearly that the high value needed to make a revolution in the period corresponding to the dominant frequency is only attained at a given measuring position for a relatively short fraction of the period. In the figure 12 the measuring point displaced from the centre of the outflow tube by close to $7.8\,\text{mm}$ and the whistling frequency was measured to be $415\,\text{Hz}$. To make one rotation at this location in the corresponding period requires an azimuthal velocity of $20.34\,\text{m/s}$, close to the maximum value in the figure. With the x-type hot wire probe in the annular outflow it was also possible to analyse the harmonic content of the velocity signal. Figure 13 shows typical spectra for the $U$ and $W$ components at a point near the maximum outward velocity position of the rotating annulus. The major component is seen as well as higher harmonics.

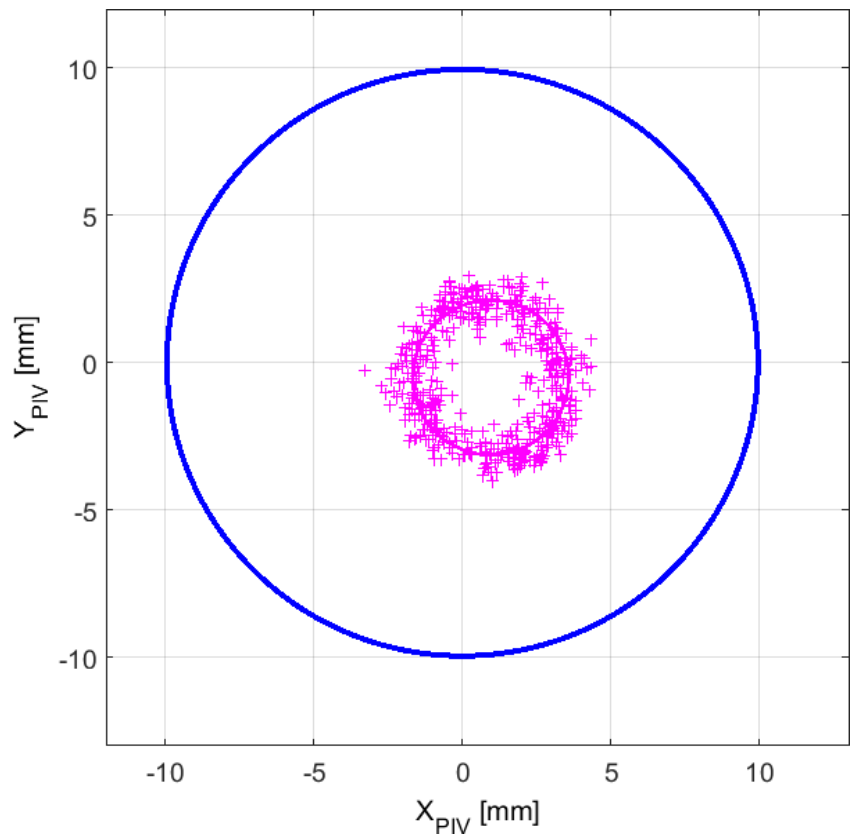

Figure 11: An assembly of vortex central points taken from PIV "snapshots" at $X = 0$.

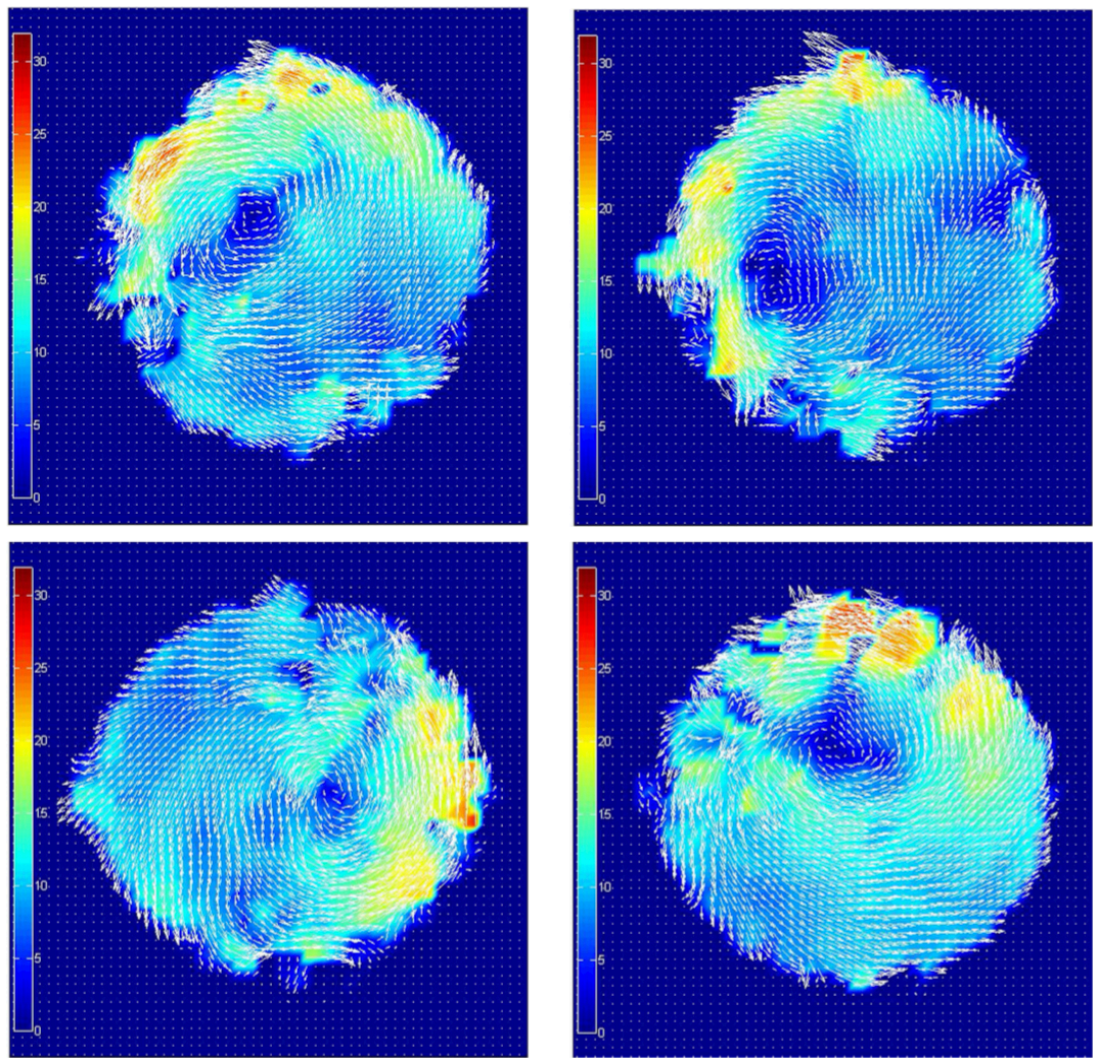

Table 3: Snapshots of movements in the exit plane

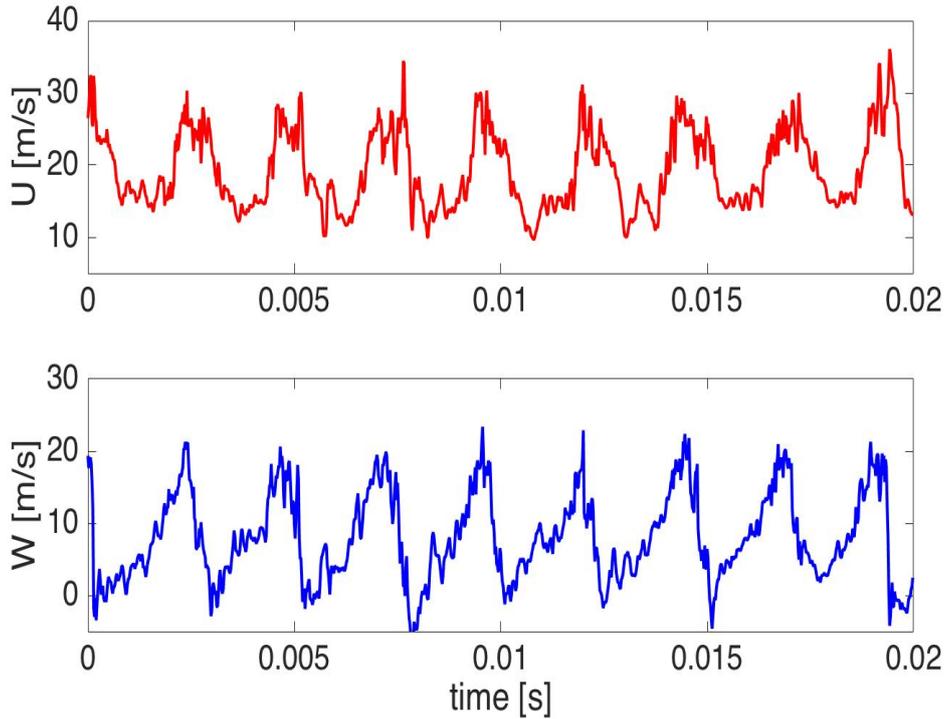

Figure 12: Instantaneous velocities in the axial $U$ (upper panel) and azimuthal $W$ (lower panel) directions (from hot-wire).

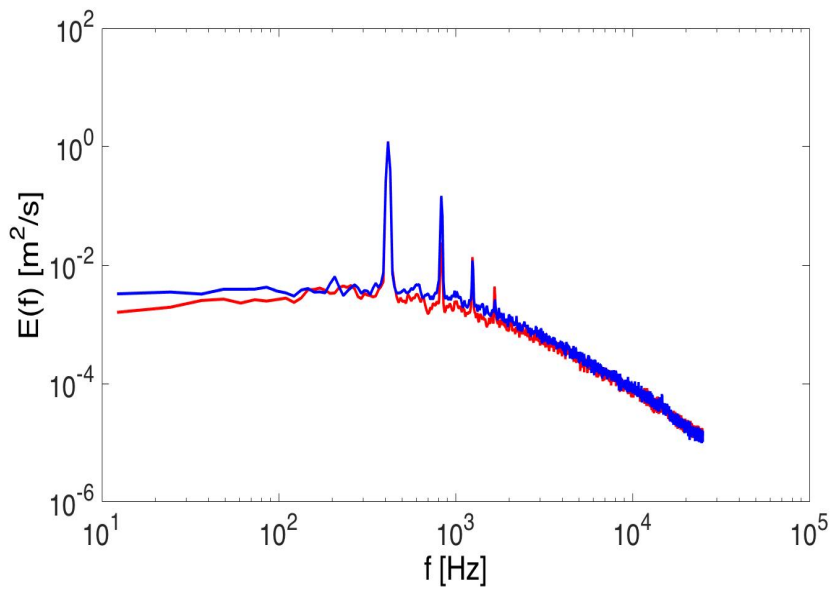

Figure 13: Velocity spectra for $U$(red) and $W$(blue) components, major peak is at 415Hz (from hot-wire).

## 4  Summary

Acoustic and flow measurements have been performed on a vortex whistle. The measurement showing that the dominant whistling frequency varies linearly with velocity agrees with earlier findings presented in the literature. The measurement on emitted acoustic power showed this quantity to vary with the frequency to the 6th power. These two measurements together indicate the source to be of the dipole type [8]. Averaged and instantaneous velocity values using PIV and hot-wire methods show the swirl of the outflow. The center of the swirl follows a nearly circular path with a concentration of the flow between the swirl center and the rim of the outflow tube. The azimuthal velocity of the flow concentration corresponds well with the velocity needed for a complete rotation in one acoustic period. This suggests that the source lies in the interaction of this concentrated flow and the rim of the outflow tube, hence constituting a rotating dipole. The flow was also observed to leave the outlet pipe in a concentric shell. An inflow along the center line in the immediate downstream region suggests that a toroidal vortex is present in this region.